\documentclass{elsart}
\usepackage{graphicx}

\newcommand{\cal}	{{}}	

\newcommand{\ALPHA}	{{A}}
\newcommand{\N}		{{\cal N}}
\newcommand{\eff}	{{\rm eff}}
\newcommand{\EW}	{{\rm EW}}
\newcommand{\BH}	{{\rm BH}}
\newcommand{\planck}	{{\rm pl}}
\newcommand{\DW}	{{\rm DW}}
\newcommand{\mini}	{{\rm min}}
\newcommand{\GUT}	{{\rm GUT}}

\newcommand{\CP}	{{\rm CP}}
\newcommand{\focus}	{{\rm fcs}}

\newcommand{\lnear}	{{\begin{array}{c} < \\[-1.3em] \sim \end{array}}}

\newcommand{\GeV}	{{\rm \;GeV}}
\newcommand{\EM}	{{\rm em}}

\newcommand{\fig}[1]	{Figure \ref{#1}}
\newcommand{\eq}[1]	{equation (\ref{#1})}
\newcommand{\EQ}[1]	{(\ref{#1})}

\begin{document}

\begin{flushright}
   hep-th/0307294
   WIS/18/03-JULY-DPP
\end{flushright}

\begin{frontmatter}

\title{%
Spherical Domain Wall formed by Field Dynamics of Hawking Radiation and\\
Spontaneous Charging-up of Black Hole}

\author{Yukinori Nagatani}
\ead{yukinori.nagatani@weizmann.ac.il}
\address{%
Department of Particle Physics,
The Weizmann Institute of Science,\\
Rehovot 76100, ISRAEL}


\begin{abstract}
 We investigate the Hawking radiation in the gauge-Higgs-Yukawa theory.
 The ballistic model is proposed as an effective description of the system.
 We find that a spherical domain wall around the black hole
 is formed by field dynamics rather than thermal phase-transition.
 The formation is a general property of the black hole
 whose Hawking temperature is equal to or greater than
 the energy scale of the theory.
 The formation of the electroweak wall and that of the GUT wall
 are shown.
 We also find
 a phenomenon of the spontaneous charging-up of the black hole by the wall.
 The Hawking radiation drives a mechanism of
 the charge-transportation into the black hole
 when C- and CP-violation are assumed.
 The mechanism can strongly transport the hyper-charge
 into a black hole of the electroweak scale.
\end{abstract}

\begin{keyword}
Hawking radiation \sep Spherical domain wall \sep Spontaneous charging-up
 
\PACS 04.70.Dy \sep 11.27.+d
\end{keyword}

\end{frontmatter}

\section{Introduction}\label{intro.sec}

Black holes radiate particles thermally
at the Hawking temperature \cite{Hawking:1975sw}.
As Schwarzschild black holes are losing their mass by the radiation,
their Hawking temperature is increasing
because the temperature is inversely proportional to the mass.
The temperature and the intensity of the radiation
increase explosively at the final stage of the evaporation,
therefore small black holes are very interesting
to particle physicists \cite{Hawking:1974rv}.

Several authors discussed that
heating-up by the Hawking radiation can thermalize the neighborhood
and thermal phase-transitions around the black holes can arise.
The QCD phase transition \cite{Cline:1996mk}
and the electroweak (EW) phase transition
\cite{Nagatani:1998gv,Nagatani:2001nz}
have been shown.
The local phase-transition around the black hole
makes the spherical domain wall
which separates the symmetric-phase-region from the broken-phase background.
Several mechanisms of the baryon-number-production by
the black holes accompanied with the EW wall have been shown and
the cosmological baryogenesis scenario by using one of the mechanisms
has been proposed
\cite{Nagatani:1998gv,Nagatani:2001nz,Nagatani:1998rt}.

The thermal phase-transition on finite volume
requires the local thermal equilibrium
which is confirmed by an enough interaction rate
in the considered volume.
It is shown that
the mean free paths of the radiated particles are much longer than
the Schwarzschild radius.
Hence the scale of the thermalized volume and
the radius of the wall should be much larger than the Schwarzschild radius.
The formation of the wall by the thermal phase transition
needs much higher Hawking-temperature than
the critical temperature of the transition
\cite{Nagatani:1998gv,Nagatani:2001nz}.
However, it is natural to ask if
the Hawking radiation whose temperature is
equal to or slightly greater than the critical temperature influences
the Higgs vacuum-expectation-value (vev) locally.

In this paper
we investigate the Hawking radiation of such a temperature
in the gauge-Higgs-Yukawa theories,
e.g., the Standard Model (SM) and the Grand Unified Theory (GUT).
We show that
a structure of the spherical domain wall around the black holes
arises by the field dynamics of the system.
The analysis seems to be difficult
because the system is fully non-equilibrium,
however,
we find that properties of the system quite simplify the analysis.

\begin{figure}
 \begin{center}
  \includegraphics{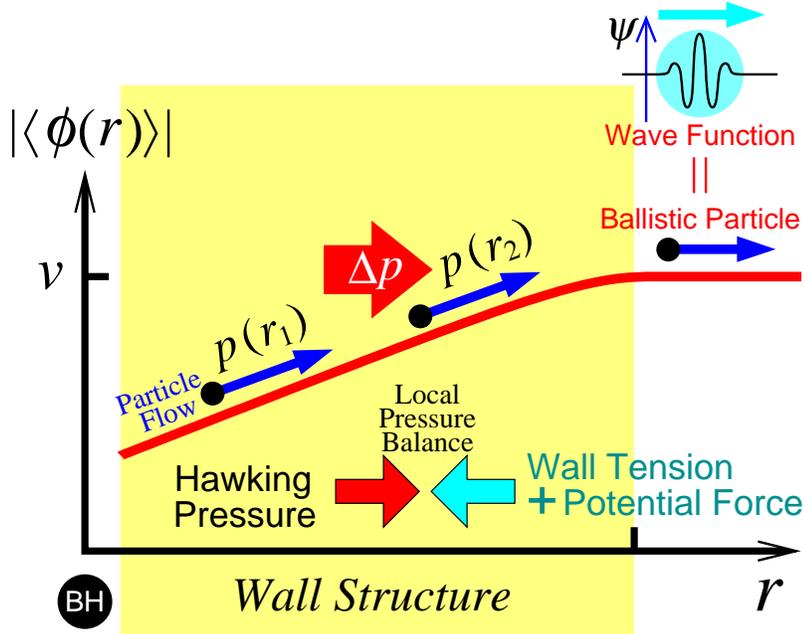}%
 \end{center}
 \caption{%
 Our basic idea for the wall-formation-mechanism by balance of pressures.
 The parameter $r$ is the distance from the center of the black hole
 and $|\left<\phi(r)\right>|$ is the absolute value of the Higgs vev.
 When once a wall-structure is formed, mass of the particle
 is depending on the distance $r$.
 A part of the momentum $\Delta p$ of the Hawking-radiated particle
 is transfered to the wall
 because the momentum is depending on the distance $r$ in the
 wall-structure: $p(r)=\sqrt{E^2 - m(r)^2}$.
 The transfered momentum causes
 a Hawking-radiation-pressure acting on the wall.
 The local balance of the pressures
 can stabilize the wall-structure.
 \label{Idea.eps}%
 }%
\end{figure}

The mean free paths of the radiated particles with gauge interactions
are much longer than the radius of the black hole
\cite{Nagatani:1998gv,Nagatani:2001nz}.
The mean wavelength of the radiated particles is approximately given by
the radius of the black hole and is shorter than the length-scale of the
wall-structure which we will consider outside the black hole.
These simplify the analysis.
The Hawking-radiated particles can be regarded as ballistic
because of their long mean-free-paths and their short wavelengths.
Their masses are proportional to the Higgs vev $|\left<\phi\right>|$.
Our basic idea of the wall-formation-mechanism is schematically
shown in the \fig{Idea.eps}.
When once a radius-depending Higgs-vev-structure $|\left<\phi(r)\right>|$,
namely the spherical wall-structure, has been formed,
the Hawking-radiated particles run rush up the slope of the structure
and they push the structure outside.
Concretely when a particle with momentum $p(r_1) = \sqrt{E^2-m(r_1)^2}$
ascends a small part of the slope from the position $r_1$ to $r_2$,
the momentum of the particle becomes
$p(r_2) = \sqrt{E^2-m(r_2)^2}$ with $m(r_2) = m(r_1) + \Delta m$.
The parameter $E$ means the conserved energy of the particle.
The part of the slope receives momentum
$\Delta p = p(r_1) - p(r_2) \simeq (m/p) \Delta m$.
Therefore there arises a pressure, named the Hawking pressure,
which is acting on the structure to expand its radius.
On the other hand the structure also feels its wall-tension and
the Higgs-potential-force,
which are acting on the structure to shrink its radius.
A balance of these pressures seems to be able to keep the structure 
stationary.
The relevant interaction in this mechanism
is only the Yukawa interaction
between a ballistic particle and the Higgs vev.

To realize the idea of the wall-formation systematically
we propose {\it the ballistic model}
as an effective description of the system.
Our model consists from
the action for the Higgs field and
that for the relativistic point particles radiated from the black hole.
We derive the effective potential
which determines the configuration of the Higgs vev around the black hole.
By considering the effective potential,
formation of the spherical domain wall around the black hole
is shown when the Hawking temperature is
equal to or greater than the energy scale of the Higgs-Yukawa theory.
We will call this mechanism
{\it the dynamical formation of the spherical wall}
in distinction from the thermal formation.
In this paper the dynamical formation of the EW wall and
that of the GUT wall are shown as concrete examples.

We also discuss a mechanism of 
{\it the spontaneous charging-up of the black hole}
as an application of the spherical wall.
The Hawking radiation is neutral for any charge, however,
the following assumptions allow us to consider the charging-up mechanism.
We assume 
(i) a CP-broken phase in the spherical wall and
(ii) a chiral charge assignment of the fermions in the field theory,
i.e., a C-violation of the theory.
These assumptions induce
a reflection-asymmetry for the charge on the wall.
Therefore
a charge-transport-mechanism from the wall into the black hole works
and it charges-up the black hole.
This mechanism is a variant application of
the charge-transport-mechanism of the electroweak baryogenesis
proposed by Cohen, Kaplan and Nelson \cite{Cohen:1991it,Cohen:1993nk}.
The assumptions are equivalent to two of the Sakharov's conditions
for the baryogenesis \cite{Sakharov:1967dj}.
When the Hawking temperature is on the EW scale,
the mechanism strongly works and 
the hyper-charge is transported into the black hole.
The black hole of the GUT temperature can also obtain several charge.


The paper is organized as follows.
In Section \ref{Formation.sec}
the formation mechanism of the spherical wall is discussed.
In Section \ref{ChargeUp.sec}
a mechanism for the spontaneous charging-up of the black hole
is discussed.
In section \ref{summary.sec}
we provide a conclusion and discussions.

\section{Dynamical Formation of Spherical Domain Wall}
\label{Formation.sec}

We consider a Hawking radiation
in the gauge-Higgs-Yukawa theory like the Standard Model (SM).
Especially
we discuss dynamics of the Higgs field
around the black hole whose Hawking temperature is
similarly equal to or greater than
the energy scale of the theory, e.g., the critical temperature.
In the situation
the mean free paths of the Hawking-radiated particles are much greater
than the length scale of the black hole \cite{Nagatani:1998gv}.
Because the mean wavelength of the radiated particle
is approximately given by the Schwarzschild radius,
the mean wavelength is similar to or smaller than
the length-scale of the domain wall which we will consider
around the black hole.
We can regard these radiated particles as ballistic.
The particles are almost freely running away from the black hole
except for the interaction with the Higgs vacuum-expectation-value (vev)
as their background.
Only the Yukawa interactions of the heavy fermions with the Higgs vev and
the gauge interactions of the heavy gauge bosons with the Higgs vev
are relevant.
These particles obtain their mass
from the Higgs vev through the interactions.
Inversely we can expect
the deformation of the Higgs vev,
i.e., the symmetry restoration or the wall formation,
by the particles with both high energy and high density
around the black hole.

The deformation can be approximately treated by a combination of
the relativistic kinematics of each radiated ballistic particle
and the dynamics of the Higgs field with the relevant interactions.
The approximated system should have Lorentz invariance
and each ballistic particle should obtain its mass by the Higgs vev.
To discuss the subject
we propose {\it the ballistic model} 
which satisfies the required conditions.
Our model is described by the action for a Higgs field $\phi(x)$
with a Higgs potential $V(\phi)$
and the action for trajectories $\{y_i^\mu(s)\}$
of relativistic point particles $\{i\}$ as the ballistic particles.
We use a metric convention $g_{\mu\nu} = {\rm diag} (+1,-1,-1,-1)$.
The action is given by\footnote{
The action for the ordinary relativistic point particle with mass $m$
is given by $-\int ds \: m \sqrt{|\dot{y}(s)|^2}$.
Our action describes point particles which obtain their mass 
by the Higgs vev.}:
\begin{eqnarray}
 S[\phi,y] &=&
  \int d^4x \; \left[ (\partial \phi)^2 - V(\phi) \right]
  \;-\;
  \sum_i \int ds_i \; Y_i \, |\phi\left(y_i(s_i)\right)| \,
         \sqrt{|\dot{y}_i(s_i)|^2},
  \label{action}
\end{eqnarray}
where $\dot{y}_i^\mu \equiv dy_i^\mu / ds_i $
and $Y_i$ is a ``extended Yukawa coupling constant''
for the point particle $i$.
The summation in the action \EQ{action} is performed over all particles.
The Higgs potential in the vacuum is given by the double-well form
\begin{eqnarray}
 V(\phi) &=&
      - \frac{1}{2} \, \mu^2 \phi^2
  \;+\; \frac{1}{2} \frac{\mu^2}{v^2} \, \phi^4,
\end{eqnarray}
which has a minimum at $|\phi| = v/\sqrt{2}$
and the constant $\mu^2 > 0$ is the square of the Higgs mass.
Mass of the particle $i$ is proportional to the Higgs vev:
\begin{eqnarray}
  m_i(x) &=& Y_i |\left<\phi(x)\right>|
  \label{mass-Yukawa-Higgs}
\end{eqnarray}
and which reproduces the relation in the ordinary gauge-Higgs-Yukawa theory.
Mass of the particle in the vacuum is given by 
$m_i = Y_i v / \sqrt{2}$.
Here we note that the summation in the action \EQ{action} is performed over
not only the fermions but also the massive gauge bosons.
Because mass of the massive gauge boson in the gauge-Higgs theory
is also proportional to the Higgs vev,
we have extended the concept of the Yukawa-coupling-constant $Y_i$
to the massive gauge boson in the action \EQ{action}\footnote{
In the simplest gauge-Higgs model with a gauge coupling constant $g$,
we have $W$-boson mass $m_W = \frac{1}{\sqrt{2}} g |\left<\phi\right>|$.
Then the extended Yukawa coupling constant is given by
$Y_W = \frac{1}{\sqrt{2}} g$.}.
Such a massive gauge boson as the ballistic particle
also obeys the mass relation \EQ{mass-Yukawa-Higgs}.
The general relativistic (GR) effect is omitted.
The reason will be discussed in section \ref{summary.sec}.

To calculate the effective potential for the Higgs field,
we fix gauge for the point particles as $s_i = x^0 = y_i^0(x^0)$.
Then we have
$\sqrt{|\dot{y}_i(x^0)|^2} = \sqrt{1 - |{\bf v}_i(x^0)|^2} \equiv
1/\gamma_i(x^0)$,
where ${\bf v}_i(x^0)$ is 3-velocity for the particle $i$.
The action becomes
\begin{eqnarray}
  S[\phi,{\bf y}] &=&
  \int d^4x
  \left[ (\partial \phi)^2 - V(\phi) 
  \;-\;
  \sum_i \; \delta^{(3)} ({\bf x} - {\bf y}_i(x^0)) 
  Y_i \, \frac{|\phi\left(x\right)|}{\gamma_i(x^0)}
  \right]
  \label{action2}.
\end{eqnarray}
By using the definition of the 4-momentum for the particle $i$:
\begin{eqnarray}
 p_{i\,\mu}
  &=& \frac{\partial L}{\partial \dot{y}_i^\mu}
  \;=\; - Y_i \: |\phi(y_i)| \frac{\dot{y}_{i\,\mu}}{\sqrt{|\dot{y_i}|^2}}
  \;=\; - Y_i \: |\phi(y_i)| \gamma_i(x^0) \; (1, \, -{\bf v}_i)\nonumber\\
  &\equiv& (-E_i, \, {\bf p}_i),
  \label{four_momentum}
\end{eqnarray}
the equation of the motion for the particle becomes
\begin{eqnarray}
  \dot{p}_{i\mu}
  \;+\;
  Y_i \frac{1}{\gamma_i(x^0)} \partial_\mu |\phi| (y_i)
  &=& 0
  \label{particle-EOM},
\end{eqnarray}
where $L$ is the integrand of the action \EQ{action2},
namely, the Lagrangian.
The parameter $E_i(x^0)$ of the particle motion means
a positive energy of the particle $i$.
The equation of the motion for the Higgs field is derived from
the gauge-fixed action \EQ{action2} and it becomes
\begin{eqnarray}
  \partial^2 \phi
   &=&
   - \frac{1}{2} \frac{\partial}{\partial\phi}
   \left[
      V(\phi)
    \;+\;
      \sum_i \; \delta^{(3)} ({\bf x} - {\bf y}_i(x^0)) 
      Y_i \, \frac{|\phi\left(x\right)|}{\gamma_i(x^0)}
   \right].
\end{eqnarray}
Here we find that the right hand side of the field-equation includes
an absolute value $|\phi|$ which is not a differentiable function.
This fact looks unnatural but this difficulty is resolvable.
By using the positive energy parameter \EQ{four_momentum},
the field equation becomes a regular form of the field:
\begin{eqnarray}
  \partial^2 \phi
    &=&
   - \frac{1}{2} \frac{\partial}{\partial\phi}
   \left[
    V(\phi)
    \;+\;
    \frac{1}{2} \sum_i \; \delta^{(3)} ({\bf x} - {\bf y}_i(x^0)) 
    Y_i^2 \, \frac{\phi^2(x)}{E_i(x^0)}
   \right].
  \label{Higgs-EOM-1}
\end{eqnarray}
If we perform the differentiation before substituting the energy parameter,
we find a Heaviside theta function $\theta(\phi) = \pm1$ temporarily
but we will obtain the same result:
$|\phi| \frac{\partial}{\partial\phi} |\phi|
= |\phi| \theta(\phi) = \phi$.
We can consider that the non-differentiability 
of the absolute-value-function $|\phi|$ in the action
is relaxed by the effect of the particle motion $\sqrt{|\dot{y}_i|^2}$.

The Higgs field is decomposed into the time-independent vev part
and the fluctuating part:
\begin{eqnarray}
  \phi(x^0,{\bf x})
  &=& \left<\phi({\bf x})\right>
  \;+\; \frac{1}{\sqrt{2}} \rho(x^0,{\bf x}).
  \label{Higgs-Decomposition}
\end{eqnarray}
The fluctuating part $\rho(x^0,{\bf x})$
can be regarded as the Higgs particles
which are propagating in the background of the position-depending Higgs vev
$\left<\phi({\bf x})\right>$.
According to our concept of the ballistic model
we consider that
the propagating mode $\rho(x^0,{\bf x})$ can
be treated as the massive particles and
be taken into the ballistic particles $\{y_i\}$.
We will precisely discuss a treatment of the ballistic Higgs particle
in the Appendix.
Mass of the Higgs propagating mode does not
obey the mass relation \EQ{mass-Yukawa-Higgs},
however,
we can show that
the action \EQ{action} correctly describes the ballistic Higgs particles
by choosing the Yukawa-coupling-constant $Y_i = \sqrt{3} \mu/v$
for the Higgs particles.
Here we redefine $\phi({\bf x})$
as the position-depending Higgs vev $\left<\phi({\bf x})\right>$ and
$\{y_i\}$ as the set of the ballistic particles
including the ballistic Higgs particles.
%
The redefined Higgs field satisfies $\partial_0 \phi = 0$.
The equation of the field-motion \EQ{Higgs-EOM-1}
becomes an effective equation of the Higgs vev by the redefinition.
The property $\partial_0 \phi = 0$ simplifies
the equation of the particle-motion \EQ{particle-EOM} as
\begin{eqnarray}
 E_i &=& Y_i \: |\phi({\bf y}_i(x^0))| \: \gamma_i(x^0)
 \label{E-const},
\end{eqnarray}
where $E_i$ becomes a constant of the particle motion.
The constant is the positive-definite energy of the particle $i$
at the rest frame.

When we put the trajectories $\{{\bf y}_i(x^0)\}$ for all particles
with their energy-constants $\{E_i\}$ into the \eq{Higgs-EOM-1},
we obtain an effective potential for the Higgs field
\begin{eqnarray}
  V_\eff(\phi) &=&
   V(\phi) 
   \;+\;
   \frac{1}{2}
   \sum_i \; \delta^{(3)} ({\bf x} - {\bf y}_i(x^0)) 
   Y_i^2 \, \frac{\phi^2({\bf x})}{E_i},
  \label{EffPot1}
\end{eqnarray}
which determines the static structure of the Higgs vev
by the field-equation
$\triangle\phi = \frac{1}{2} \frac{\partial}{\partial\phi} V_\eff$.
Here we adopt a differential particle number-density
$dE \times \N_{f}(E;{\bf x})$ for particle species $f$.
The effective potential for the Higgs field can be written down as
\begin{eqnarray}
  V_\eff(\phi) &=&
   V(\phi) 
   \;+\;
   \frac{1}{2}
   \phi^2
   \sum_f Y_f^2
   \int \frac{dE}{E} \N_f(E;{\bf x}),
  \label{EffPot2}
\end{eqnarray}
where the summation in the \eq{EffPot2} is performed over
all particle-species in the theory.
Then the effective potential should be
\begin{eqnarray}
 V_\eff(\phi, {\bf x}) &=&
    +   \frac{1}{2} \, \mu_\eff^2({\bf x})  \, \phi^2
  \;+\; \frac{1}{2} \frac{\mu^2}{v^2} \, \phi^4,
 \label{EffPot3}
\end{eqnarray}
where we have defined the effective $\mu^2$-parameter as
\begin{eqnarray}
 \mu_\eff^2({\bf x})
  &=& -\mu^2
  \;+\; \sum_f Y_f^2 \int \frac{dE}{E} \N_f(E;{\bf x}).
  \label{mu_eff}
\end{eqnarray}

Next we consider that the particle distribution $dE \times \N_f(E;{\bf x})$
is produced by the Hawking radiation from the Schwarzschild black hole
with mass $m_\BH$.
By the Hawking process
all kinds of the particles in the field theory are radiated
from the horizon whose radius is given by the Schwarzschild radius
$r_\BH = \frac{1}{4\pi}\frac{1}{T_\BH}$.
The radiated particles have thermal spectrum at the Hawking temperature
$T_\BH = \frac{1}{8\pi}\frac{m_\planck^2}{m_\BH}$.
The particle distribution around the black hole is
approximately given by
\begin{eqnarray}
 dE \; \N_f
  &=& \frac{1}{4} \frac{g_f}{(2\pi)^3} \; f_{T_\BH}(E)
  \; 4\pi E^2 dE  \;\times\; \left(\frac{r_\BH}{r}\right)^2,
  \label{distribution}
\end{eqnarray}
where
\begin{eqnarray}
 f_{T_\BH}(E) &:=& \frac{1}{e^{E/T_\BH} \pm 1}
\end{eqnarray}
is the Bose-Einstein or the Fermi-Dirac distribution function
of the temperature $T_\BH$
and $g_f$ means a degree of freedom for the particle-species $f$.
The leading contribution from the Hawking radiation
is given by the distribution \EQ{distribution}.
Later we will discuss the formation of the wall by
the Hawking-radiated particles.
The wall influences the trajectories of the particles and
slightly deforms the distribution of the particles.
In the distribution \EQ{distribution}
we have ignored this backreaction.
The ignored contribution 
can be evaluated by solving the equation of the particle-motion
$|\dot{\bf y}_i|^2 = 1 - (Y_i \phi({\bf y}_i)/E_i)^2$
derived from the \eq{E-const}
and it will be a future subject.

\begin{figure}
 \begin{center}
  \includegraphics{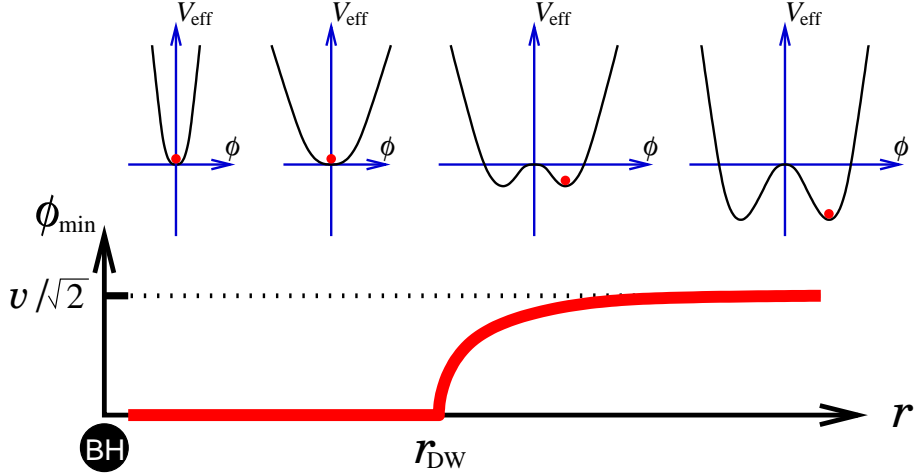}%
 \end{center}
 \caption{%
 The distribution of the effective Higgs potential $V_\eff(\phi,r)$
 around a black hole.
 The parameter $r$ is the distance from the center of the black hole
 and $r_\DW=\ALPHA/\mu$ is the radius
 where the sign of $\mu_\eff^2(r)$ changes.
 The thick curve indicates the value of the Higgs field $\phi_\mini(r)$
 which minimizes the effective Higgs potential at each point $r$.
 Since the value $v/\sqrt{2}$ is
 the ordinary Higgs vacuum expectation value,
 a boundary condition $\phi_\mini(r\rightarrow\infty) = v/\sqrt{2}$
 is required.
 \label{WallConcept.eps}%
 }%
\end{figure}

By substituting the distribution \EQ{distribution}
into the effective Higgs mass \EQ{mu_eff}
we obtain
\begin{eqnarray}
 \mu_\eff^2(r)
  &=& -\mu^2
  \;+\; \frac{\ALPHA^2}{r^2},
\end{eqnarray}
where we have defined a dimensionless ``wall-formation-constant''
\begin{eqnarray}
 \ALPHA^2 &\equiv& \frac{1}{768\pi^2} \sum_f \tilde{g}_f Y_f^2 
  \label{ALPHA}
\end{eqnarray}
which is depending on the field theory
and we have defined an effective $g_f$ as
\begin{eqnarray}
 \tilde{g}_f &=&
  \left\{
   \begin{array}{c@{\hspace{8mm}}l}
    g_f & (f:{\rm boson})\\
    \frac{1}{2}g_f & (f:{\rm fermion})
   \end{array}
  \right..
\end{eqnarray}
Here we find that the sign of $\mu_\eff^2(r)$ changes
at the ``characteristic wall-radius''
\begin{eqnarray}
 r_\DW &\equiv& \frac{\ALPHA}{\mu}
 \label{rDW}
\end{eqnarray}
as
$\mu_\eff^2(r)$ is negative for $r>r_\DW$ and is positive for $r<r_\DW$.
The distribution of the Higgs effective-potential $V(\phi,r)$
around the black hole is shown in the \fig{WallConcept.eps}.
The effective potential is minimized at each position $r$ by
\begin{eqnarray}
 \phi_\mini(r) &=&
  \left\{
   \begin{array}{c@{\hspace{8mm}}l}
    \displaystyle
     \frac{v}{\sqrt{2}}
     \left[ \; {1 - \left(\frac{r_\DW}{r}\right)^2} \; \right]^{1/2}
     & (r > r_\DW) \\[5mm]
     0 & (r \leq r_\DW)
   \end{array}
  \right..
  \label{phi_min}
\end{eqnarray}
Therefore there arises a local-vacuum depending on $r$ continuously
and we can expect the formation of
the spherical domain wall of the radius $r_\DW$
when $r_\DW > r_\BH$.
The vacuum continuously varying with the dependence on the position
is one of the distinctive features in the non-equilibrium system.
%

\begin{figure}[htbp]%
 \begin{center}
  \includegraphics[scale=0.85]{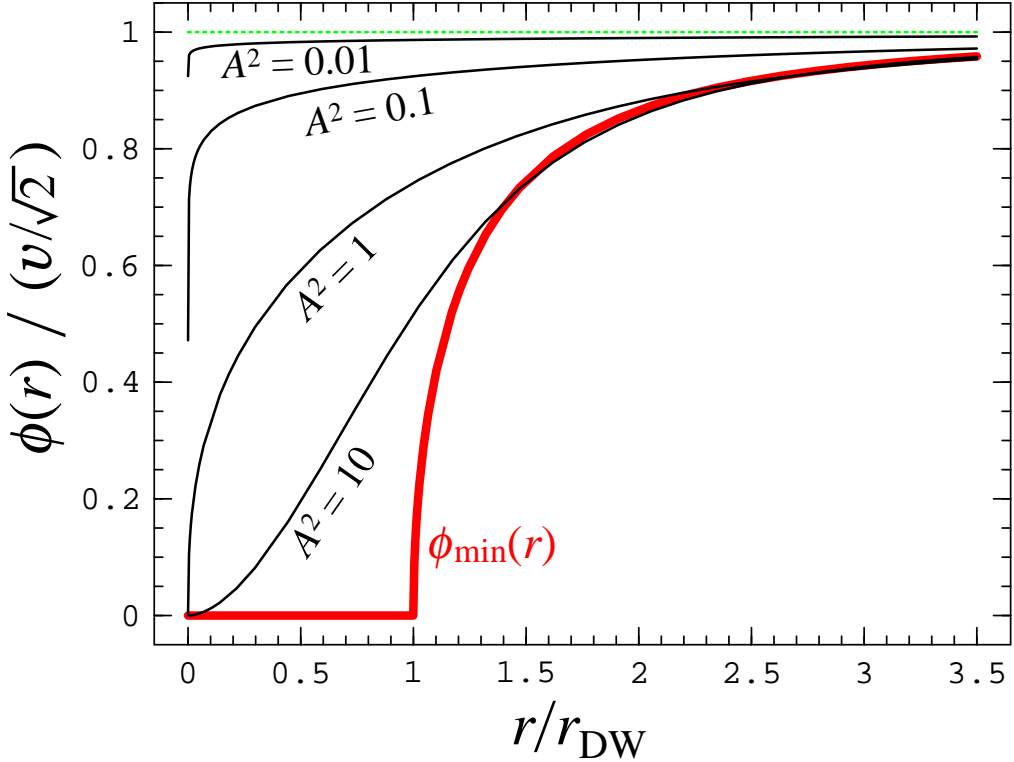}%
 \end{center}
 \caption{%
 The profiles of the spherical-wall around a black hole.
 The parameter $r$ is distance from the center of the black hole
 and $r_\DW$ is a characteristic-wall-radius.
 The thin curves $\phi(r)$ are resultant Higgs vevs
 depending on the wall-formation-constant $\ALPHA^2$.
 They are spherically-symmetric stationary solutions of
 the effective field-equation \EQ{Higgs-EOM-2}
 which associates with the effective potential $V_\eff(\phi,r)$
 in \EQ{EffPot3}.
 The thick curve $\phi_\mini(r)$ indicates the value of the Higgs field
 which minimizes the effective Higgs potential at each point.
 The dotted line means the ordinary Higgs vev
 $\left<\phi\right> = v/\sqrt{2}$.
 \label{WallForm.eps}%
 }%
\end{figure}

The form of the Higgs vev around a black hole,
namely, the structure of the wall,
is calculated as a spherically-symmetric stationary solution of the
motion-equation for the Higgs field with the effective potential \EQ{EffPot3}:
\begin{eqnarray}
 \triangle\phi(r)
 &=&
 \frac{1}{2}\frac{\partial}{\partial\phi} V_\eff(\phi,r),
  \label{Higgs-EOM-2}
\end{eqnarray}
where a boundary condition $\phi(r\rightarrow\infty) = v/\sqrt{2}$
is required.
The \eq{Higgs-EOM-2} can be solved numerically and
the result is shown in the \fig{WallForm.eps}.
The form of the Higgs vev is depending on
the wall-formation-constant $\ALPHA^2$ defined in \EQ{ALPHA}.
Generally speaking,
the curve of the Higgs vev $\phi(r)$ approaches $\phi_\mini(r)$
when $\ALPHA^2$ becomes large.
Here we find that
the Hawking radiation can form the spherical wall
or wall-like structure around the black hole for the non-zero $\ALPHA^2$
and the radius of the spherical wall is characterized
by the parameter $r_\DW$.

The condition for the wall-formation becomes
\begin{eqnarray}
  r_\BH &<& r_\DW,
  \label{condition}
\end{eqnarray}
because the Schwarzschild radius $r_\BH$
should be smaller than the characteristic-wall-radius $r_\DW$.
This condition can be rewrite as
\begin{eqnarray}
 T_\BH &>& T_\BH^*,
\end{eqnarray}
where we have defined the critical Hawking-temperature
\begin{eqnarray}
 T_\BH^* &\equiv& \frac{\mu}{4\pi A}
\end{eqnarray}
for the dynamical wall-formation.
It should be noticed that the characteristic-wall-radius $r_\DW$ in \EQ{rDW}
is not depending on the Hawking temperature.
Once the wall is formed,
the shape of the wall hardly changes
in spite of the change of the Hawking temperature
during the evaporation of the black hole.
The integrated particle-number-flux from the black hole
which is derived from \EQ{distribution}
is proportional to the Hawking temperature $T_\BH$.
On the other hand,
the effect of the ballistic particles upon the Higgs vev
is inversely proportional to the Hawking temperature
because of the form of the effective potential \EQ{EffPot1}
with the relation $E \sim T_\BH$.
Therefore the influence of the Hawking radiation on the wall
is not depending on the Hawking temperature.

Now we can discuss the formation of the wall
in the realistic field theories, e.g.,
the electroweak (EW) theory or
the grand unified theory (GUT).
In the EW theory
the heavy particles which contribute to create the wall
are the top quarks, the weak bosons $Z^0$, $W^{\pm}$ and the Higgs bosons.
Approximately we have the wall-formation-constant for the EW wall:
\begin{eqnarray}
 \ALPHA_\EW^2 &\equiv& \frac{1}{768\pi^2} \sum_f \tilde{g}_f Y_f^2 \nonumber\\
 &=& 10^{-3} \times
     \left[ 0.979 + 0.065 \left(\frac{\mu_\EW}{100\GeV}\right)^2 \right]
  \;\simeq\; 1/960 \sim 1/800
  \label{ALPHAEW}
\end{eqnarray}
which is depending on the Higgs mass $\mu_\EW = 100 \sim 200 \GeV$.
The critical Hawking-temperature for the dynamical formation of
the EW wall becomes
\begin{eqnarray}
 T_\BH^{*\EW} &\equiv& \frac{\mu_\EW}{4\pi\ALPHA_\EW}
  \;\sim\; 2 \, \mu_\EW.
\end{eqnarray}
Consequently
the Hawking radiation produces the spherical EW domain wall
whose structure is given in the \fig{WallForm.eps}
with $\ALPHA^2 = \ALPHA_\EW^2 \simeq 0.001$
when the Hawking temperature of the black hole
is similar to or greater than the EW scale.
The characteristic radius of the wall is given by
\begin{eqnarray}
 r_\DW^\EW &=& \ALPHA_\EW/\mu_\EW \;\simeq\; 0.03/\mu_\EW
\end{eqnarray}
In the $SU(5)$-GUT
there are, at least, lepto-quarks $X,Y$ and $\bf 24$-Higgs $\Phi$
as heavy particles of the GUT scale mass $\mu_\GUT \simeq 10^{16}\GeV$.
The same analysis results in
\begin{eqnarray}
 \ALPHA_\GUT^2 &\simeq& 1/200,\\
 T_\BH^{*\GUT} &\simeq& \mu_\GUT,\\
 r_\DW^\GUT    &\simeq& 0.07/\mu_\GUT.
\end{eqnarray}
Then the spherical GUT wall is also formed
when the Hawking temperature is greater than the GUT scale.
Finally we can conclude that
the spherical wall around a black hole is dynamically formed
in most of the field theories
which have a Higgs mechanism
when the Hawking temperature is greater
than the energy scale of the field theory.

\section{Spontaneous Charging-up of the Black Hole}
\label{ChargeUp.sec}

We discuss that the Hawking radiation can charge up the black hole
by the effect of the spherical wall discussed in the previous section.
We need two assumptions such that
(i) the domain wall has a CP-broken phase and
(ii) the field theory has fermions with chiral charge-assignment.
The Standard Model (SM) satisfies the second assumption
because the left-handed quark and the right-handed quark have 
different hyper-charges.
The SM with the two Higgs extension (2HSM)
or the Minimal Supersymmetric SM (MSSM) can satisfy the first assumption.
When the first assumption is satisfied,
the reflection-rate on the wall of the left-handed fermions
is different from that of the right-handed fermions.
The domain wall has a charge-reflection-asymmetry
when both assumptions are satisfied.
The Hawking radiation is charge-neutral, however,
the black hole obtains net charge 
because
the reflected particle-flux is not charge-neutral due to the asymmetry
and a part of the reflected particles return into the black hole.
Consequently the net charge is transported into the black hole.
This process is similar to the
``charge transport mechanism on the thin wall regime''
in the electroweak baryogenesis
proposed by Cohen, Kaplan and Nelson \cite{Cohen:1991it,Cohen:1993nk}.
In the charge-transport-mechanism of the electroweak baryogenesis,
the hyper-charge is transported
from the thin EW wall to the symmetric-phase-region
and it boosts up the baryon-number-creation
by the sphaleron process.
On the other hand
our mechanism of the charge-transportation charges up the black hole.

The rate for the charge-transportation into the black hole is given by
\begin{eqnarray}
 \frac{dQ}{dt} = \sigma_\BH \times C_\focus F_Q,
\end{eqnarray}
where
$\sigma_\BH$ is the cross section for the absorption into the black hole,
$F_Q$ is the reflected charge flux on the wall and
the dimensionless parameter $C_\focus$ is a focusing factor.
We adopt the horizon area as the absorption-cross-section:
\begin{eqnarray}
  \sigma_\BH &=& 4\pi r_\BH^2.
  \label{absorption}
\end{eqnarray}
Strictly speaking, the absorption-cross-section $\sigma_\BH$
is slightly different from the horizon area \EQ{absorption}
with a factor of $O(1)$
and is depending on the energy of the particle $E$
due to the general relativistic effect.
The absorption-cross-section for the high-energy limit (the eikonal limit)
is given by $\sigma_\BH(E\rightarrow\infty) = \frac{27\pi}{4} r_\BH^2$.
The spectrum of the Hawking radiation is slightly different
from the thermal distribution
because the flux of the Hawking radiation
is proportional to the $\sigma_\BH(E)$,
which is known as the gray-body effect.
The absorption-cross-section \EQ{absorption}
is consistent with the flux of the Hawking radiation
given in the distribution \EQ{distribution}.

In the spherical reflector
any particles radiated from a neighborhood of the center of the reflector
return into the same neighborhood when they are reflected.
Then the incoming flux to the horizon from the spherical wall
is greater than the reflected flux $F_Q$ on the wall.
This is a focusing effect by the spherical reflector.
The flux on the horizon can be written down as $C_\focus F_Q$
by the focusing factor $C_\focus \geq 1$.
We have $(r_\BH/r_\DW)^2 \times C_\focus = 1$ when the effect maximally works
and we have $C_\focus = 1$ in the absence of the effect.

The charge flux on the wall is given by
\begin{eqnarray}
 F_Q &=& \sum_{f \in \chi} \;\;
  \int_{m_f}^{\infty} dE \; {\cal N}_f(E ; r_\DW) \;
  \Delta Q_f \;
  \Delta R_f(E).
  \label{Q-Flux}
\end{eqnarray}
The summation
is performed over all species $\chi$ of the chiral-charged fermions,
where the particle-species $f \in \chi$ does not distinguish both
the chirality of the particle and the particle/anti-particle.
The number $\Delta Q_f \equiv Q_{f_L} - Q_{f_R}$
means the difference 
between the charge of the left-handed fermion $f_L$ and
that of the right-handed fermion $f_R$.
This number is related to the C-violation for the theory.
The function
\begin{eqnarray}
 \Delta R_f(E) &\equiv&
  R_{f_R \rightarrow f_L}(E) - R_{\bar{f}_R \rightarrow \bar{f}_L}(E)
\end{eqnarray}
describes the difference between the reflection-probabilities,
where
the reflection-probability $R_{f_R \rightarrow f_L}(E)$ is
a probability which the right-handed fermion $f_R$
is reflected to the left-handed fermion $f_L$
and
the reflection-probability $R_{\bar{f}_R \rightarrow \bar{f}_L}(E)$ is
that for the anti-fermion: $\bar{f}_R \rightarrow \bar{f}_L$.
These probabilities are depending on the particle-energy $E$.
The non-zero $\Delta R_f(E)$ is related with the CP-violation of the wall.
When we put the structure of the CP-broken phase in the wall,
we can numerically evaluate $\Delta R_f(E)$
according to the method proposed by Cohen, Kaplan and Nelson
\cite{Cohen:1991it}.

The wall-solution (see the \fig{WallForm.eps})
shows the absolute value of the Higgs vev $|\left<\phi(r)\right>|$,
namely, the wall profile without a complex phase.
To evaluate the reflection-asymmetry $\Delta R_f(E)$
we assume the structure of the CP-broken phase in the wall:
\begin{eqnarray}
 \left<\phi(r)\right>
  &=&
  f(r) \times \exp
  \left[
   i \: \Delta\theta_\CP \left( 1 - \frac{f(r)}{v/\sqrt{2}} \right)
  \right], \label{CP-profile}
\end{eqnarray}
where $\Delta\theta_\CP$ is the amount of the CP-broken phase in the wall
and the function $f(r) \equiv |\left<\phi(r)\right>|$ is
a profile of the Higgs vev in the wall.
The wall-structure \EQ{CP-profile} is defined for $r>r_\BH$.
The profile-function $f(r)$ should be
a solution of the effective-field-equation \EQ{Higgs-EOM-2}.
The complex-phased wall-structure \EQ{CP-profile}
is the same as that in the EW baryogenesis \cite{Cohen:1991it,Cohen:1993nk}
except for the profile function\footnote{
In the EW baryogenesis
$f(z) = [1 + \tanh(z/z_\DW)]/2$ is assumed
as the profile function of the wall formed
by the first order phase transition,
where $z_\DW$ is a thickness of the wall.
}.
The reflection-asymmetry $\Delta R_f(E)$ of the wall is depending on
the CP-broken phase $\Delta\theta_\CP$,
the wall-formation-constant $A^2$,
the energy scale of the field-theory $\mu$,
the energy of the fermion $E$ and
mass of the fermion $m_f$.

\begin{figure}[htbp]%
 \includegraphics[scale=0.85]{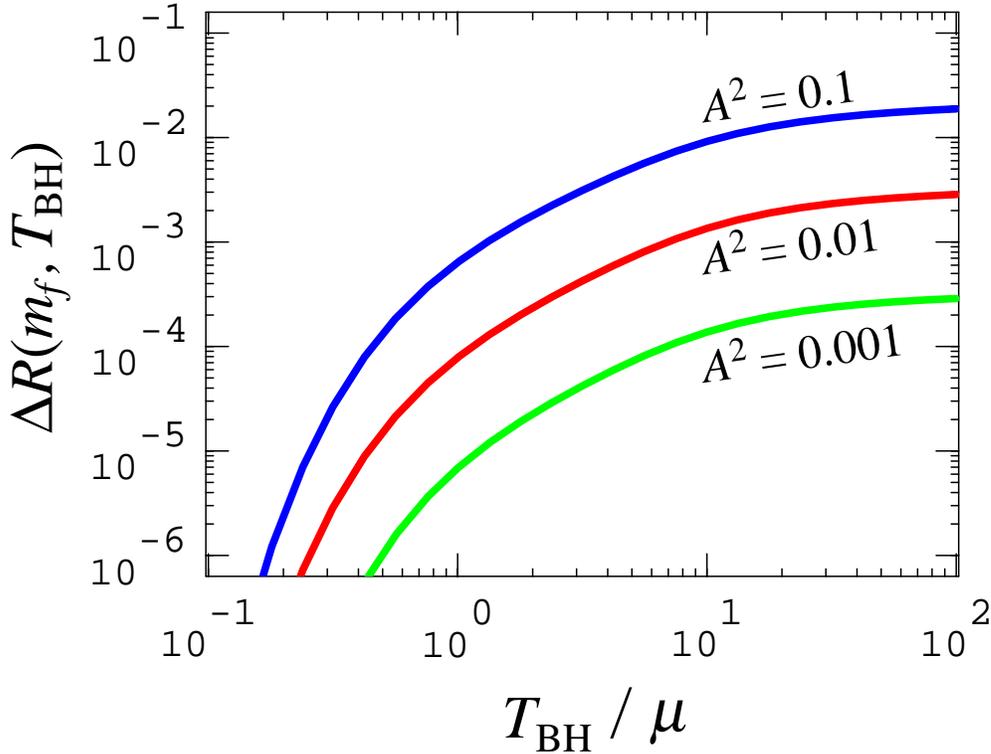}%
 \caption{%
 Numerical results of the 
 dimensionless integrated reflection-asymmetry $\Delta{\cal R}(m_f,T_\BH)$
 for the wall given in the \eq{CP-profile} with
 wall-formation-constants
 $\ALPHA^2=0.1$, $\ALPHA^2=0.01$ and $\ALPHA^2=0.001$
 and with a maximum CP-violated phase $\Delta\theta_\CP = \pi$.
 We have assumed $\mu = m_f$ for simplicity.
 \label{DR.eps}%
 }%
\end{figure}

\begin{figure}[htbp]%
 \includegraphics[scale=0.85]{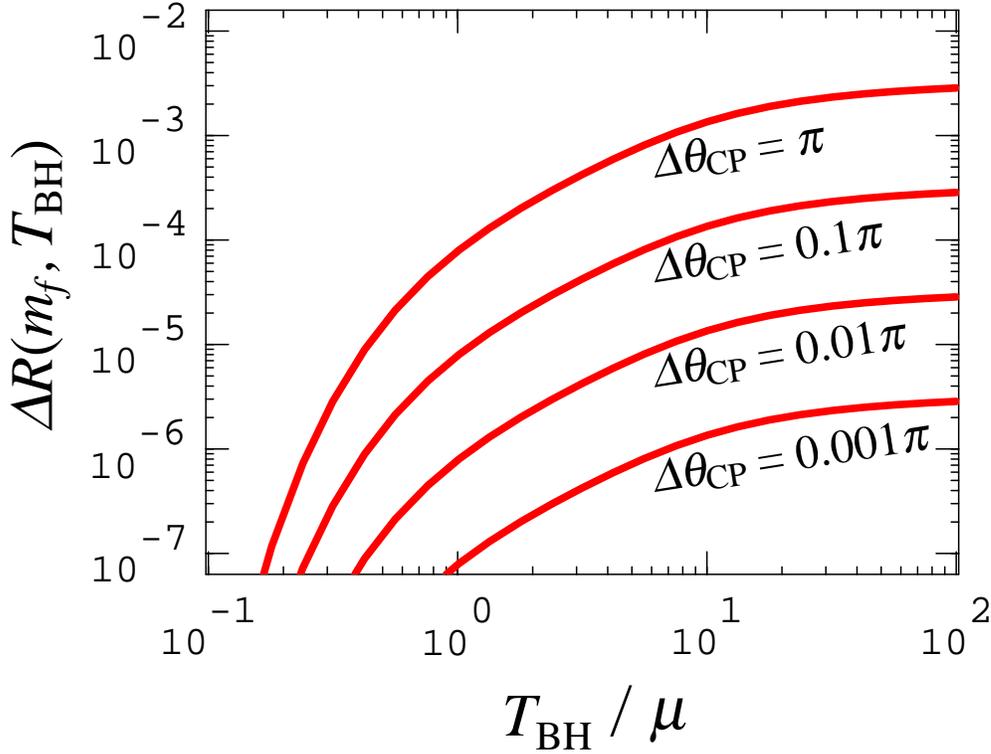}%
 \caption{%
 Numerical results of $\Delta{\cal R}(m_f,T_\BH)$
 with a typical wall-formation-constant $\ALPHA^2=0.01$
 and with several CP-violated phases
 $\Delta\theta_\CP/\pi=1,\ 0.1,\ 0.01,$ and $ 0.001$.
 \label{DRP.eps}%
 }%
\end{figure}

By substituting the particle-distribution \EQ{distribution},
the charge flux \EQ{Q-Flux} becomes
\begin{eqnarray}
 F_Q
  &=&
  \left(\frac{r_\BH}{r_\DW}\right)^2 C_\focus
  \,\times\,
  \sum_{f \in \chi} \;
  g_f \, \Delta Q_f \, m_f^3 \;
  \Delta{\cal R}(m_f,T_\BH),
\end{eqnarray}
where we have defined a reflection-asymmetry
integrated over the thermal distribution
\begin{eqnarray}
 \Delta{\cal R}(m_f,T_\BH)
  &=&
  \frac{1}{8\pi^2}
  \int_1^\infty d\xi \;
  \xi^2 \; f_{T_\BH}(m_f \xi) \;
  \Delta R_f(m_f \xi).
\end{eqnarray}
We have numerically calculated $\Delta R_f(E)$
and we show the resultant reflection-asymmetry
$\Delta {\cal R}_f(m_f,T_\BH)$ in the \fig{DR.eps} and \fig{DRP.eps}.
In these calculation we have assumed $\mu = m_f$ for simplicity.

The black hole of the Hawking temperature $T_\BH$
has a finite lifetime
$\tau_\BH = \frac{20}{\pi^2 g_*} \frac{m_\planck^2}{T_\BH^3}$,
where $g_*$ is the total degree of freedom
with a fermion correction\footnote{
The definition is $g_* \equiv g_{\rm b} + (7/8) g_{\rm f}$,
where $g_{\rm b}$ is the total degree of the freedom of bosons and
$g_{\rm b}$ is that of the fermions.
In the SM with the gravity, we have $g_* = 108.75$.}.
The total charge transported into the black hole in its lifetime
with assuming no mechanism of charge-loss works
is useful for evaluating the strength of the mechanism.
In section \ref{summary.sec}
we will discuss several mechanisms of the charge-loss.
The total transported charge becomes
\begin{eqnarray}
 Q
 &=&
  \int^{\tau_\BH} dt \frac{dQ}{dt} \nonumber\\
 &=&
  \frac{15}{\pi^3 g_*} m_\planck^2
  \sum_{f \in \chi} g_f \Delta Q_f m_f^3
  \int_{T_\BH}^\infty \frac{dT}{T^6}
  \Delta{\cal R}(m_f,T), \label{charge}
\end{eqnarray}
where we have assumed the maximum focusing effect.
The integration in the \eq{charge} can be performed
by using the numerical form of $\Delta{\cal R}(m_f,T)$.
According to the numerical results in the \fig{DR.eps} and the \fig{DRP.eps},
the function $\Delta{\cal R}(m_f,T)$ 
has meaningful value for $T_\BH > \mu$ and
is exponentially dumping for $T_\BH < \mu$.
Therefore we approximately obtain
the total charge transported into the black hole
whose initial Hawking temperature $T_\BH$ 
is similar to or smaller than $\mu$:
\begin{eqnarray}
 Q
 &\simeq&
  \frac{3}{\pi^3} \,
  \Delta{\cal R}(\mu,\mu)
  \left(\frac{m_\planck}{\mu}\right)^2
  \sum_{f \in {\rm H}\chi}
  \frac{g_f}{g_*}
  \Delta Q_f,
\end{eqnarray}
where we have put a simplification $m_f = \mu$
for all related fermions $f$, i.e.,
the summation
is performed over only species of the heavy fermions ${\rm H}\chi$.
Our numerical analysis results in
$\Delta{\cal R}(\mu,\mu) \simeq 10^{-4} \times \Delta\theta_\CP/\pi$
for $\ALPHA^2=1/100$
as the EW wall or as the GUT wall,
then we obtain
\begin{eqnarray}
 Q &\simeq& 
  10^{-5}
  \times
  \frac{\Delta\theta_\CP}{\pi}
  \left(\frac{m_\planck}{\mu}\right)^2
  \sum_{f \in {\rm H}\chi}
  \frac{g_f}{g_*}
  \Delta Q_f.
\end{eqnarray}
Finally we conclude that
the spherical wall can charge up the black hole non-trivially
when the energy scale of the wall satisfies
$\mu \lnear 10^{-3} \times m_\planck \simeq 10^{16}\GeV$.

When we consider a black hole whose Hawking temperature is
on the EW scale $\mu\simeq100\GeV$,
the spherical EW wall arises and
the hyper-charge $Q_Y$ is transported into the black hole.
Mainly the top quark, $g_f = 3$ and $\Delta Q_f = 1/2$,
carries the hyper-charge into the black hole.
The total transported hyper-charge becomes
$Q_Y \simeq 10^{25} \times \Delta\theta_\CP$.
The transported hyper-charge can become meaningful number
even if $\Delta\theta_\CP$ is small,
then we conclude that
the spontaneous charging-up mechanism of the black hole works strongly.
We expect that the mechanism also works by
the CP-asymmetry of the wall derived by 
the Kobayashi-Maskawa phase of the SM.
When a black hole of the GUT temperature is considered,
several charge can be transported into the black hole.
The amount of the charge is
depending on the chiral charge assignment $\Delta Q_f$ of the GUT.

\section{CONCLUSION AND DISCUSSIONS}\label{summary.sec}

%
In this paper
the Hawking radiation in the gauge-Higgs-Yukawa theory is investigated
by the effective model
based on the ballistic property of the Hawking-radiated particles.
The dynamical formation of the spherical domain wall
around the black hole is shown.
Our model, named the ballistic model, is described by 
the ordinary Higgs action with a double well potential
and the action for the many relativistic point particles
coupled to the Higgs field.
The ballistic model quite simplifies
the analysis of the Higgs vev structure around the black hole
as compared with directly analyzing the field-equations 
for the gauge-Higgs-Yukawa theory.

The ballistic model is valid when
(i) the mean free path of the Hawking-radiated particle
is greater than the wall radius $r_\DW$ and
(ii) the mean wavelength $\lambda$ of the particle is smaller than $r_\DW$.
The mean wavelength $\lambda$ is approximately given by
the Schwarzschild radius $r_\BH$.
The radius of the wall we have concluded in \EQ{condition} satisfies
$r_\DW > \lambda \sim r_\BH$,
then our analysis is consistent with the second condition.
If the background of the black hole is not thermalized,
the first condition is always satisfied.
When the Hawking temperature $T_\BH$ is so much high
that the Hawking radiation thermalizes its neighborhood
and the plasma-temperature in the neighborhood locally exceeds
the critical temperature of the phase transition,
the thermal formation of the spherical wall arises
\cite{Nagatani:1998gv,Nagatani:2001nz}
instead of the dynamical formation.
The transition from the dynamical wall to the thermal wall
may be interesting subject.

Transitional time-evolutions of the vev,
like the process of the initial wall-formation,
are computed by \EQ{Higgs-EOM-2}
with adding a kinetic term $\partial_t^2 \phi$
in a sense of the time-depending mean-field.
The time-scale of the wall-formation
is estimated to be $\tau_\DW \simeq \mu^{-1}$ because
both the wall-radius \EQ{rDW} and the effective potential \EQ{EffPot3}
have the same energy scale $\mu$.
The lifetime of the black hole is
$\tau_\BH = \frac{20}{\pi^2 g_*} \frac{m_\planck^2}{T_\BH^3}$.
When we consider a black hole of $T_\BH \simeq \mu$,
the ratio becomes
$\tau_\DW / \tau_\BH \simeq g_*^{-1} (m_\planck/T_\BH)^2$.
The ratio for the GUT wall is $10^3 \sim 10^4$.
Therefore there is enough lifetime of the black hole for the wall-formation
when energy-scale of the phase-transition
is equal to or lower than the GUT scale.

General relativistic (GR) effects are omitted in our analysis.
The black hole has been assumed as an particle-source
with a thermal spectrum of the Hawking temperature.
No reliable treatment for the GR corrections
to the Hawking-radiated particles has been known.
For example, a blue-shift effect near the horizon arises
when we consider that
the trajectory of the radiated particles obeys the Schwarzschild metric.
The effect implies that
the energy of the radiated particles near horizon is
much higher than the Hawking temperature.
No matter how low the Hawking-temperature is,
the effect allows us to consider the formation of 
the spherical wall near the horizon dynamically or thermally.
This effect is interesting, however,
has difficulties to treat properly.
The problem of the GR-effect will be the future subject.

We have evaluated the influence from the radiated particles
into the Higgs vev and have found the formation of the wall.
On the other hand,
there should arise
the backreaction from the wall into the motion of the particles.
The largest backreaction is the reflection
of the low-energy particles by the wall.
When $T_\BH \gg m_f$, the backreaction is ignored
because most of the radiated particle pass through the wall.
In the case
the form of the wall is insignificantly deformed by the backreaction and
the Higgs vev in the wall becomes slightly smaller.
When $T_\BH \sim m_f$, the backreaction cannot be ignored
because many of the radiated particles are reflected on the wall.
We will subsequently develop the argument of the backreaction.

One may consider that our solution for the Higgs field
shown in the \fig{WallForm.eps}
is not a domain wall
according to the usual terminology ---
a domain wall is a region of a false vacuum
which separates into different real vacua.
However our terminology ``domain wall''
is suitable because of the following reasons.
The local vacuum $\phi_\mini(r)$ in \EQ{phi_min}
is continuously varying (see the \fig{WallConcept.eps})
because the effective potential $V_\eff(\phi,{\bf x})$ in \EQ{EffPot3}
is apparently depending on the position $\bf x$.
The continuous vacuum is roughly divided into two vacua, namely,
the symmetry-restored local-vacuum
$\phi_\mini=0$ for the neighborhood of the black hole as $r < r_\DW$
and the symmetry-broken vacuum $\phi_\mini=v/\sqrt{2}$ for $r \gg r_\DW$.
Our solution separates these vacua.
Our solution $\phi(r)$ does not correspond
to the continuous vacuum $\phi_\mini(r)$ for $r_\BH < r \lnear r_\DW$
(see the \fig{WallForm.eps}),
i.e., the region can be considered as a false vacuum.
Consequently the properties of our solution agree with
the ordinary definition of the domain wall.

We have also discussed the mechanism of the spontaneous charging up
of the black hole by the effect of the spherical domain wall.
The mechanism works
when a C-violation of the field theory and
a CP-broken phase in the wall are assumed.
Our analysis results that the mechanism works
when the initial Hawking temperature of the black hole
is smaller than about $10^{16}\GeV$.
The black hole of EW size strongly drives the mechanism
and the hyper-charge is transported into the black hole by the EW wall.
The black hole of the GUT size can obtain several charge by the GUT wall.

On the other hand,
a mechanism for the spontaneous charge-loss of the black hole
has been discussed.
Gibbons first proposed this subject
by the semi-classical method \cite{Gibbons:1975kk,Hiscock:1990ex}
and recently Gabriel discussed that this result can be confirmed
by the functional method \cite{Gabriel:2000mg}.
Their works were briefly as follows.
The strong electric field around the black hole induces
the pair creation of the charged particles.
One of the pair-created particles falls into the black hole
and the other runs away from the black hole.
The outgoing particles carry away the charges from the black hole.
Consequently they concluded that
a kind of Schwinger process could work for discharging the black hole.
These calculation are reliable
when the radius of the black hole is greater
than the Compton wave length of the electron, i.e., $r_\BH > 1/m_e$.
Because the radius of the black hole discussed in this paper
is smaller than the EW length-scale,
we can not directly apply their process of the charge-loss
to our system.
Several authors discussed the subject
for the charge loss of the smaller black hole $r_\BH < 1/m_e$
\cite{Page:1977um,Khriplovich:1998si},
however, we do not know a relevant treatment in this case.

There is not only the quantum process of the charge-loss
but also the classical mechanism preventing the black hole charging-up.
The radiated particles whose charge is opposite to the black hole
are attracted to the black hole.
When the electric charge is transported and
the charge of the black hole becomes $Q$,
the particles of charge $q$ feel the electrostatic potential as
$V(r) = \alpha_\EM Q q/ r$.
If we assume that the black hole is only a particle-source of
energy $E \simeq T_\BH \simeq 1/r_\BH$,
the charging-up by our mechanism stops
when the charge of the black hole becomes
$Q = 1/\alpha_\EM \simeq 137$
because
the radiated particle of the opposite charge
can not across the potential barrier.
In this paper we have used the same assumption,
then this charge-bound may be natural.

The subjects of the charge-up and the charge-loss
are related to the problem of the remnant;
``after the completion of the Hawking radiation,
whether something is left or not?'' and
``if something is left, what is it?''.
One of the candidates for the remnant is the extremal-charged black,
which is quantum-mechanically stable.
If the process of the charge-loss does not work,
the extremal-charged black hole is left as the remnant
because
the black hole looses its mass by the Hawking radiation
and approaches the extremal-charged hole.
Zel'dovich discussed that
a black hole leaves a remnant with a Planck mass scale
\cite{Zeldovich:1984}.
If the charge-loss process works and later the process stops
as the charge becomes small,
a Planck-massive remnant with several (hyper-)charges
is left after the end of the Hawking radiation.
There arises no singularity in the solution of the Dirac equation
around a electric point charge $Q$ if $Q < 1/\alpha_\EM \simeq 137$.
The solution with a singularity indicates
the fall of the charged particle into the center,
then we may expect the stop of the charge-loss.

Such a remnant can be applicable to explain the dark matter
of the universe \cite{Nagatani:2001nz}.
The scenario is as follows.
Primordial black holes are created by the density-fluctuation 
in the very early universe
and are later made to dominate over the universe by the universe expansion.
The Hawking radiation of the black hole reheats up the universe
and
the spherical EW wall thermally formed around the black hole
provides the baryon number for the universe.
The black hole obtains some charge by our charge-up process
and becomes the Planck-massive remnant
as the main constituent of the dark matter.
Finally we expect that
the formation of the spherical wall around the black hole
plays crucial roles in the cosmology.

\begin{flushleft}
 {\bf ACKNOWLEDGMENTS}
\end{flushleft}

 I would like to thank
 Ofer~Aharony, Micha~Berkooz, Alex~Buchel,
 Hikaru~Kawai, Kazunori~Kohri, Barak~Kol,
 Masao~Ninomiya, Mihoko~Nojiri and Joan~Simon
 for useful discussions.
 I am grateful to Kei~Shigetomi
 for helpful advice and also for careful reading of the manuscript.
 The work has been supported in part by
 the Grant-in-Aid for Scientific Research
 from the Ministry of Education, Culture, Sports, Science and Technology
 (JSPS-1999-03665)
 and also has been supported by the Koshland Postdoctoral Fellowship of
 the Weizmann Institute of Science.

\section*{APPENDIX: Ballistic Model for Higgs Particle}\label{Appendix.sec}

In this appendix
we discuss a treatment of the Higgs particle as the ballistic particle.
The action for the Higgs field $\phi$ is given by
\begin{eqnarray}
 S[\phi] &=& \int d^4x \left[ (\partial\phi)^2 - V_\eff(\phi) \right]
\end{eqnarray}
with the effective potential
\begin{eqnarray}
 V_\eff(\phi) &=&
    \frac{1}{2} \mu^2_\eff({\bf x}) \phi^2
  + \frac{1}{2} \frac{\mu^2}{v^2} \phi^4.
\end{eqnarray}
The Higgs field is decomposed into
\begin{eqnarray}
 \phi(x^0,{\bf x})
  &=& \phi_s({\bf x}) + \frac{1}{\sqrt{2}} \rho(x^0,{\bf x}),
  \label{Higgs-Decomposition-ap}
\end{eqnarray}
where
$\phi_s({\bf x}) \equiv \left<\phi({\bf x})\right>$ is a
time-independent Higgs vev
and $\rho(x^0,{\bf x})$ is a massive Higgs propagating mode
in the background of the Higgs vev.
When we consider the ordinary vacuum given by
the potential with $-\mu^2_\eff(x) = \mu^2 > 0$,
we obtain a constant Higgs vev $\phi_s({\bf x}) = v/\sqrt{2}$
and a Higgs mass $m_\rho = \mu$.

The Higgs vev $\phi_s({\bf x})$ is determined by the solution of
the field equation with $\partial_0 \phi_s = 0$:
\begin{eqnarray}
 0 &=&
  \triangle \phi_s
  - \frac{1}{2}
  \left.\frac{\partial V_\eff}{\partial\phi}\right|_{\phi=\phi_s}.
  \label{EOM-phi_s}
\end{eqnarray}
The equation of the motion
for the Higgs propagating mode $\rho(x^0,{\bf x})$ becomes
\begin{eqnarray}
 \frac{1}{\sqrt{2}} \left[ \partial_0^2 \rho - \triangle \rho \right]
  &=&
    \triangle \phi_s 
  - \frac{1}{2}
    \left.\frac{\partial V_\eff}{\partial\phi}
    \right|_{\phi=\phi_s + \frac{1}{\sqrt{2}}\rho}\nonumber\\
  &=&
    \triangle \phi_s 
  - \frac{1}{2}
    \left.\frac{\partial V_\eff}{\partial\phi}\right|_{\phi=\phi_s}
  - \frac{1}{2}
    \left.\frac{\partial^2 V_\eff}{\partial\phi^2}\right|_{\phi=\phi_s}
    \frac{\rho}{\sqrt{2}}
  + \cdots.
  \label{EOM-rho}
\end{eqnarray}
In the last deformation in \EQ{EOM-rho}
we have expanded the right hand side by $\rho/\sqrt{2}$ around $\phi_s$.
By substituting \EQ{EOM-phi_s} into \EQ{EOM-rho} we obtain
\begin{eqnarray}
 \partial_0^2 \rho - \triangle \rho 
  &=&
  - \frac{1}{2}
    \left.\frac{\partial^2 V_\eff}{\partial\phi^2}\right|_{\phi=\phi_s}
    \rho
  + \cdots,
\end{eqnarray}
therefore, we find that mass of the Higgs propagating mode is given by
\begin{eqnarray}
 m^2_\rho(\mu^2, \phi_s^2) &\equiv&
  \frac{1}{2}
  \left.\frac{\partial^2 V_\eff}{\partial\phi^2}\right|_{\phi=\phi_s}
  = \frac{1}{2} \mu_\eff^2 + 3 \frac{\mu^2}{v^2} \phi_s^2
  \label{Higgs-Mass}
\end{eqnarray}
and is not proportional to the Higgs vev.

\begin{figure}
 \begin{center}
  \includegraphics{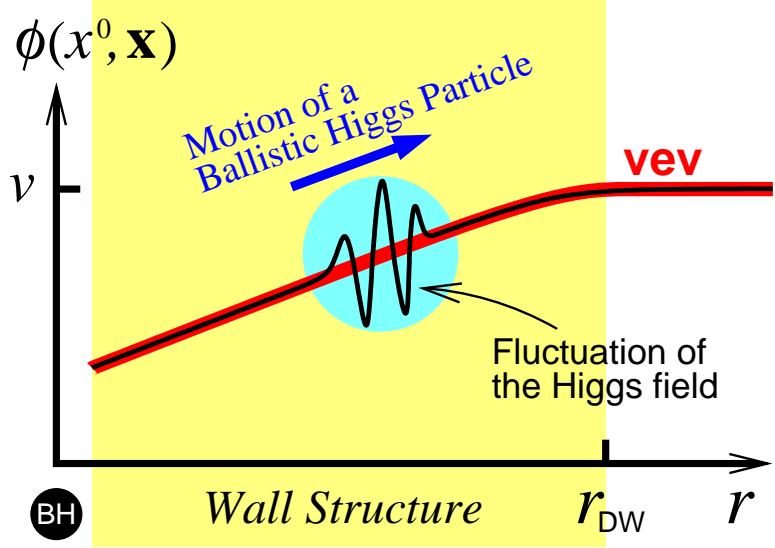}%
 \end{center}
 \caption{%
 A schematic picture of the ballistic Higgs particle.
 The thick curve is the Higgs vev $\phi_s(r) \equiv \left<\phi(r)\right>$
 which is the time-independent spherically-symmetric solution of \EQ{EOM-phi_s}
 and thin curve is the Higgs field $\phi(x^0,{\bf x})$.
 The fluctuation $\rho(x^0,{\bf x})$ of the Higgs field in
 \EQ{Higgs-Decomposition-ap}
 is regarded as a ballistic Higgs particle propagating
 in the background of the Higgs vev.
 \label{BallisticHiggs.eps}%
 }%
\end{figure}

The mass relation for the Higgs particle becomes \EQ{Higgs-Mass}
rather than the linear relation \EQ{mass-Yukawa-Higgs}
derived from the action \EQ{action}.
We should consider a new ballistic model
which includes the Higgs particles as the ballistic particles
and derives the mass relation \EQ{Higgs-Mass}.
Our concept of the ballistic Higgs particle
is schematically shown in the \fig{BallisticHiggs.eps}.
The model is described by the action for the Higgs vev $\phi_s({\bf x})$
and the action for the trajectories $\{y_i\}$
of the ballistic Higgs particles $\{i\}$:
\begin{eqnarray}
 S[\phi_s,y]
  &=&
  \int d^4x \; \left[ (\partial \phi_s)^2 - V_\eff(\phi_s) \right]\nonumber\\
  &&
  -\sum_{i} \int ds_i \;
  m_\rho(\mu^2({\bf y}_i), \phi_s^2({\bf y}_i)) \;
  \sqrt{|\dot{y}_i(s_i)|^2},
  \label{action-A1}
\end{eqnarray}
where the summation is performed over all Higgs particles
around the black hole.
The action \EQ{action-A1} is obtained by
changing the factor $Y_i |\phi(y_i)|$ in the action \EQ{action}
into $m_\rho(\mu^2({\bf y}_i), \phi_s^2({\bf y}_i))$.
The field $\phi_s({\bf x})$ does not include the propagating mode,
then the condition $\partial_0 \phi_s = 0$ on the field equation
is required.
The parameter $\mu_\eff^2({\bf x})$ in the effective potential $V_\eff(\phi_s)$
has been determined by evaluating the effect of all the ballistic particles
around the black hole except for the Higgs particles.
We will consider the additional effect on $V_\eff(\phi_s)$
by the ballistic Higgs particles.
The equation of the motion for the ballistic Higgs particle becomes
\begin{eqnarray}
 E_i = m_\rho(\mu^2({\bf y}_i), \phi_s^2({\bf y}_i)) \, \gamma_i(x^0).
  \label{Higgs-Energy}
\end{eqnarray}
The equation for Higgs vev $\phi_s$ becomes
\begin{eqnarray}
 0 &=& \triangle\phi_s
  - \frac{1}{2} \frac{\partial V_\eff}{\partial\phi_s}
  - \frac{1}{2} \sum_i \delta^{(3)}({\bf x}-{\bf y}_i) \frac{1}{\gamma_i}
    \frac{\partial m_\rho}{\partial\phi_s}.
  \label{Higgs-EOM-A1}
\end{eqnarray}
The factor in \EQ{Higgs-EOM-A1} is computed by using \EQ{Higgs-Energy}:
\begin{eqnarray}
 \frac{1}{\gamma_i} \frac{\partial m_\rho}{\partial\phi_s}
  &=& \frac{m_\rho}{E} \frac{\partial m_\rho}{\partial\phi_s}
  =
  \frac{1}{2} \frac{1}{E}
  \frac{\partial}{\partial\phi_s} m_\rho^2
  =
  \frac{\partial}{\partial\phi_s}
  \left(
   \frac{3}{2} \frac{1}{E} \frac{\mu^2}{v^2} \phi_s^2
  \right).
\end{eqnarray}
The \EQ{Higgs-EOM-A1} becomes
\begin{eqnarray}
 0 &=& \triangle\phi_s
  - \frac{1}{2}\frac{\partial}{\partial\phi_s}
   \tilde{V}_{\eff}(\phi_s),
  \label{Higgs-EOM-A2}
\end{eqnarray}
where we have defined a new effective potential
\begin{eqnarray}
 \tilde{V}_{\eff}(\phi_s)
  &=&
   V_\eff(\phi_s)
   + \frac{1}{2} \sum_i \delta^{(3)}({\bf x}-{\bf y}_i(x^0)) \;
     3 \frac{\mu^2}{v^2} \; \frac{\phi_s^2({\bf x})}{E_i}
  \label{EffPot-A1}
\end{eqnarray}
including the effect of the ballistic Higgs particles.
By comparing the effective potential \EQ{EffPot1} and \EQ{EffPot-A1},
the difference between the ordinary ballistic particle
and the ballistic Higgs particle is only constant factor.
These effective potentials are identical by defining
the Yukawa-coupling-constant for the ballistic Higgs particle as
\begin{eqnarray}
 Y_\rho = \sqrt{3} \frac{\mu}{v}.
  \label{Yukawa-For-Higgs}
\end{eqnarray}
Both the action \EQ{action}
with the Yukawa-coupling-constant \EQ{Yukawa-For-Higgs}
for the Higgs particles and the action \EQ{action-A1}
for the ballistic Higgs particles
result the same effective potential for the wall-formation.
Finally we conclude that
the action \EQ{action} also describes the ballistic Higgs particles
by choosing the Yukawa-coupling-constant \EQ{Yukawa-For-Higgs}
for the Higgs particles.


\begin{thebibliography}{99}

\bibitem{Hawking:1975sw}
S.~W.~Hawking,
Commun.\ Math.\ Phys.\  {\bf 43}, (1975) 199.

\bibitem{Hawking:1974rv}
S.~W.~Hawking,
Nature\ {\bf 248} (1974), 30.

\bibitem{Cline:1996mk}
D.~B.~Cline,
Nucl.\ Phys.\ A\ {\bf 610}, (1996) 500C.

\bibitem{Nagatani:1998gv}
Y.~Nagatani,
Phys.\ Rev.\ D\ {\bf 59}, (1999) 041301.

\bibitem{Nagatani:2001nz}
Y.~Nagatani,
hep-ph/0104160.

\bibitem{Nagatani:1998rt}
Y.~Nagatani,
hep-ph/9805455.

\bibitem{Cohen:1991it}
A.~G.~Cohen, D.~B.~Kaplan and A.~E.~Nelson,
Nucl.\ Phys.\ {\bf B349}, (1991) 727;

\bibitem{Cohen:1993nk}
A.~G.~Cohen, D.~B.~Kaplan and A.~E.~Nelson,
Ann.\ Rev.\ Nucl.\ Part.\ Sci.\  {\bf 43}, (1993) 27.

\bibitem{Sakharov:1967dj}
A.~D.~Sakharov,
Pisma Zh.\ Eksp.\ Teor.\ Fiz.\ {\bf 5}, (1967) 32
[JETP Lett.\  {\bf 5}, (1967) 24].

\bibitem{Gibbons:1975kk}
G.~W.~Gibbons,
Commun.\ Math.\ Phys.\  {\bf 44}, (1975) 245;

\bibitem{Hiscock:1990ex}
W.~A.~Hiscock and L.~D.~Weems,
Phys.\ Rev.\ D\ {\bf 41}, (1990) 1142.

\bibitem{Gabriel:2000mg}
C.~Gabriel,
Phys.\ Rev.\ D\ {\bf 63}, (2001) 024010.

\bibitem{Page:1977um}
D.~N.~Page,
Phys.\ Rev.\ D\ {\bf 16}, (1977) 2402.

\bibitem{Khriplovich:1998si}
I.~B.~Khriplovich,
gr-qc/9812060.

\bibitem{Zeldovich:1984}
Ya.~B.~Zel'dovich,
in {\it Quantum Gravity},
Proceedings of the Second Seminar,
Moscow, USSR, 1981,
edited by M.~A.~Markov and P.~C.~West
(Plenum, New York, 1984).

\end{thebibliography}

\end{document}